\documentclass[%
superscriptaddress,
nofootinbib,
amsmath,amssymb,
]{revtex4-1}

\usepackage{bm, mathtools}
\usepackage{amsmath ,amssymb}
\usepackage{graphicx}
\usepackage{graphics,epsfig}
\usepackage{dcolumn}
\usepackage[latin1]{inputenc}
\usepackage[T1]{fontenc}
\usepackage{enumitem}
\usepackage{physics}
\usepackage{adforn}
\usepackage{type1cm}        
\usepackage{makeidx}         
\usepackage[bottom]{footmisc}
\usepackage{newtxtext}       %
\usepackage{newtxmath}       

\newcommand{\Eqref}[1]{Eq.\ \eqref{#1}}

\newcommand{\ie}{i.e}

\newcommand{\eg}{e.g}
\newcommand{\etc}{etc}
\newcommand{\ee}{\mathrm e}
\newcommand{\im}{\mathrm i}
\newcommand{\del}{\partial}
\newcommand{\kB}{k_{\rm B}}

\makeatletter
\newcommand*{\mt}{%
	{\mathpalette\@mt{}}%
}
\newcommand*{\@mt}[2]{%
	\raisebox{\depth}{$\m@th#1\intercal$}%
}
\makeatother

\newcommand{\Cemmath}[4]{
  \ifthenelse{\equal{#1}{inline}}
             {$#2$#4}
             {\begin{#1}#2 \,#4\label{#3}\end{#1}}%
}

\renewcommand{\phi}{\varphi}
\newcommand{\xv}{\vb*{x}}
\newcommand{\UU}{V}
\newcommand{\CC}{\vb{C}}
\newcommand{\md}{\rho}

\newcommand{\qr}{q}
\newcommand{\qvc}{\vb*{q}_{\!\rm c}}
\newcommand{\CO}{\vb{\Omega}}
\newcommand{\co}{\Omega}
\newcommand{\DD}{\vb{D}}
\newcommand{\tm}{t_{\rm m}}
\newcommand{\te}{t_{\rm e}}
\newcommand{\Scomp}{S_{\rm c}}
\newcommand{\Sint}{\tilde{S}}
\newcommand{\Sbar}{\bar{S}}
\newcommand{\gr}{g}
\newcommand{\gv}{\vb*{\gr}}
\newcommand{\erfi}{\erf{\!\mathrm i}}
\newcommand{\nhat}{\hat{\vb*{n}}}
\newcommand{\uhat}{\hat{\vb*{u}}}
\newcommand{\vhat}{\hat{\vb*{v}}}
\newcommand{\TS}{\vb{T}_{\circ}}
\newcommand{\ts}{T^{\circ}}
\newcommand{\tss}{T_{\circ}}
\newcommand{\TC}{\vb{T}_{\times}}
\newcommand{\tc}{T^{\times}}
\newcommand{\tcc}{T_{\times}}
\newcommand{\MM}{\vb{M}}
\newcommand{\mm}{m}
\newcommand{\NN}{\tilde{\vb{M}}}
\newcommand{\nn}{\tilde{m}}
\newcommand{\mmm}{\mm^{-}}
\newcommand{\nnn}{\nn^{-}}
\newcommand{\nnp}{\nn^{+}}

\begin{document}

\title{Magnetic resonance assessment of effective confinement
	anisotropy with orientationally-averaged single and double diffusion
	encoding}

\author{Cem Yolcu}
\affiliation{Dept.\ of Biomedical Engineering, Link\"oping University, Sweden}
\author{Magnus Herberthson}%
\affiliation{Dept.\ of Mathematics, Link\"oping University, Sweden}
\author{Carl-Fredrik Westin}
\affiliation{Dept.\ of Radiology, Brigham and Women's Hospital, Harvard Medical School, Boston, MA}
\author{Evren \"Ozarslan}
\affiliation{Dept.\ of Biomedical Engineering, Link\"oping University, Sweden}
\affiliation{Center for Medical Image Science and Visualization, Link\"oping University, Sweden.}


\begin{abstract}
Porous or biological materials comprise a multitude of micro-domains
containing water. Diffusion-weighted magnetic resonance measurements
are sensitive to the anisotropy of the thermal motion of such
water. This anisotropy can be due to the domain shape, as well as the
(lack of) dispersion in their orientations. Averaging over
measurements that span all orientations is a trick to suppress the
latter, thereby untangling it from the influence of the domains'
anisotropy on the signal. Here, we consider domains whose anisotropy
is modeled as being the result of a Hookean (spring) force, which has
the advantage of having a Gaussian diffusion propagator while still
retaining the fact of finite spatial range for the diffusing
particles. Analytical expressions for the powder-averaged signal under
this assumption are given for so-called single and double diffusion
encoding schemes, which sensitize the MR signal to the diffusive
displacement of particles in, respectively, one or two consecutive time
intervals.
\end{abstract}

\maketitle


\section{Introduction}

Magnetic resonance has proved to be an extremely effective tool to
peer into materials and tissues noninvasively. It manipulates the
magnetic orientation of molecules pervading the material into a
natural precession which emits radio waves. By imposing a
spatially-varying magnetic field (hence frequency of precession), the
emitted radio frequency signal is made to encode in its spectrum the
coordinates of the molecules that emit it. This way, the signal can be
spectrally decomposed to trace back what proportion of it originates
from where, that is, from which `voxel.' Commonly achieved voxel
sizes are in the neighborhood of a millimeter.

Another use of a spatially-varying precession rate involves sensitizing the
signal to the motion of the molecules; diffusion in particular. When
the molecules trace out random (Brownian) paths where different
locations they visit impart different precessional angles on them,
their precessions lose coherence, attenuating the sum of their emitted
radio waves. This attenuation of MR signal, specifically its response
to the direction of the gradient in precession rate, can reveal or
quantify how mobile the molecules are along different directions. The
difference of mobility can arise from pore boundaries, impurities,
cell membranes, \etc. Hence the diffusion-attenuated signal encodes
the influence of structures that can be significantly smaller than
voxel dimensions. Tracing out axon bundles in human brain white matter
is for instance a widely employed application of this principle. This modality of
magnetic resonance imaging, which we refer to as diffusion MR, is the
subject of this contribution.

While the anisotropy (\ie., variance under a rotation transformation)
of a single pore or a cell may be easily visualized, one would be
mistaken to make a one-to-one connection with that and the anisotropy
of the signal (\ie., the response of the signal to orientations of the
specimen or the apparatus). For instance, the signal of a voxel
consisting of an unaligned mixture of cylindrical aqueous compartments
will be less sensitive to rotations than one consisting of an aligned
bundle of cylinders. The anisotropy of the individual compartments is common in the two
examples, but the aligned case has more ensemble anisotropy.

In some sense, then, ensemble anisotropy confounds compartment
anisotropy at the signal level, and eliminating it pronounces features
at the subvoxel level. One way to achieve this is to take an average
of the signal over all orientations. If the material allows it, it can
be ground into a powder to that effect; hence the term \emph{powder}
averaging. However this is generally impossible in bio-medical
applications. Then, repeated applications of a measurement protocol in
different orientations is the avenue to follow.

\begin{figure}[t!]
  \centering\includegraphics[]{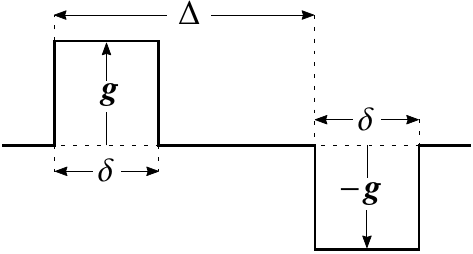}
  \caption{The pulse sequence for single diffusion encoding, with time
    running along the horizontal direction.}\label{fig:sde}
\end{figure}

In this contribution, we are concerned with two particular diffusion
MR schemes, single and double diffusion encoding (SDE and DDE), orientationally averaged in
the aforementioned fashion to eliminate ensemble anisotropy. The
single encoding scheme \cite{StejskalTanner65} is the bread and butter of most
applications, employing a magnetic field gradient that remains on for
a specified duration in one direction, and then the opposite direction
after a specified delay (Fig.\ \ref{fig:sde}). The signal then encodes
the probability of Brownian displacement between the application of
the two pulses. 

The double diffusion encoding (DDE) scheme employs two single-encoding blocks in succession \cite{Cory90b}, in different directions in general, before the signal is read (Fig.\ \ref{fig:dde}). This method has been studied extensively in recent years mostly because it allows the anisotropy of the microdomains to be quantified \cite{ChengCory99}, which has important implications for medical imaging.  The reader is referred to the reviews \cite{Callaghanbook2,Finsterbusch2011review,Shemesh12,Shemesh16nomenclature,Novikov19NBMreview} for in-depth presentation of the method. In a nutshell, the DDE technique encodes
into the signal the joint probability of two Brownian displacements
taking place between the two pulses of each block. For freely
diffusing molecules, the two displacements are uncorrelated
\cite{RiskenBook}. However, when restrictions, and arguably
inhomogeneities and forces, are present, this is no longer the case
\cite{Mitra95}, hence imparting signatures of compartment size onto
the signal. As mentioned above,  DDE employing gradient
blocks in different directions is sensitive to anisotropy of
subdomains inside the voxel \cite{ChengCory99, Callaghan02}, but not
independently of ensemble anisotropy \cite{OzarslanJCP08}.

While we do not consider restricted diffusion in the strict
sense of restriction by hard walls in this article, we do respect the
finite range of motion of the molecules by the aid of a harmonic
attractive force \cite{Uhlenbeck30,Callaghan80,LeDoussal92,MitraHalperin95,Yolcu16}. This is an \emph{effective} model
which can mimic restrictive walls with reasonably tractable
mathematics, and is actually approached when the gradient pulses are
long \cite{Ozarslan17FiP}.

This chapter is organized as follows. We first present the signal
arising from an effectively confined domain under double diffusion
encoding. Afterwards we derive analytical expressions for its powder
average. We then treat the case of single diffusion encoding as an
extreme case of double encoding and give its corresponding powder
average expressions. The special cases of ``stick'' and ``pancake''
geometry are compared to their counterparts for free diffusion. The
article is concluded after discussions based on the results of the
previous sections.

\section{Double diffusion encoding at the compartment level}

Here we derive the double-encoded diffusion MR signal arising from a
compartment which is characterized by an effective spring force
attracting the molecules toward the center. Such a force mimics the
confining effect of walls, membranes, \etc., with minimal mathematical
burden.

The spring (Hookean) force influences the motion of the molecules via
a quadratic potential\footnote{In units of the thermal energy scale $\kB T$, where $\kB$ is the Boltzmann constant and $T$ is the absolute temperature.}
\begin{align}
  \UU (\xv) = \tfrac12 \xv^{\mt} \CC \xv \,,
\end{align}
defining the \emph{confinement tensor} $\CC$, which can be taken to
have $\CC^{\mt}=\CC$ without loss of generality. However, in order for
the steady state molecule number density $p_{\rm st} (\xv) \propto
\ee^{-\UU(\xv)}$ to be normalizable, $\CC$ must be positive
(semi)definite.\footnote{Vanishing confinement (\ie., free diffusion)
  has an unnormalizable steady state, but it can be handled.} Under
this (or any) potential, the magnetization density $\md (\xv,t)$ evolves according to
\begin{align} \label{eq:Hooke}
  \del_t \md (\xv,t) = (\DD \nabla) \cdot \ee^{-\UU(\xv)}
  \nabla \ee^{\UU(\xv)} \md (\xv,t) - \im \gv (t) \cdot \xv \md(\xv,t) \,.
\end{align}
Here, we have assumed that diffusion is governed by a spatially-uniform, possibly anisotropic
diffusivity tensor $\DD$ and that diffusion encoding is achieved by a
magnetic field gradient waveform $\gv(t)$. Note that we absorb the
gyromagnetic ratio $\gamma$ into $\gv(t)$ so that it has dimensions of
time$^{-1}\,$length$^{-1}$.\footnote{Alternatively, $\gv(t)$ can be
  called a precession rate gradient waveform.}

The signal $\Scomp = \int \dd[3]x\, \md(\xv,t)$ arising from a single
such confined compartment under a general encoding waveform $\gv(t)$
can be found thanks to the Brownian paths having a Gaussian
probability measure under the potential \eqref{eq:Hooke}
\cite{Yolcu16}:
\begin{align}
  \Scomp = \exp\!\left(
  - \tfrac12 \qvc^{\mt}\!(0) \DD \CO^{-1} \qvc(0)
  - {\int_0^{\te}} \!\!\dd{\tau} \qvc^{\mt}\!(\tau) \DD \qvc(\tau)
  \right) \,, \label{eq:signal.conf}
\end{align}
with the generalized encoding wave vector
\begin{align} \label{eq:q}
  \qvc(t) = {\int_{t}^{\te}} \!\!\dd{\tau} \ee^{-\CO (\tau-t)} \gv (\tau) \,.
\end{align}
Here, $\te$ is the duration of the encoding protocol, and
\begin{align}
  \CO = \DD \CC \label{eq:Omega}
\end{align}
is a matrix of equilibration rates.

\begin{figure}[!t]
  \centering\includegraphics[]{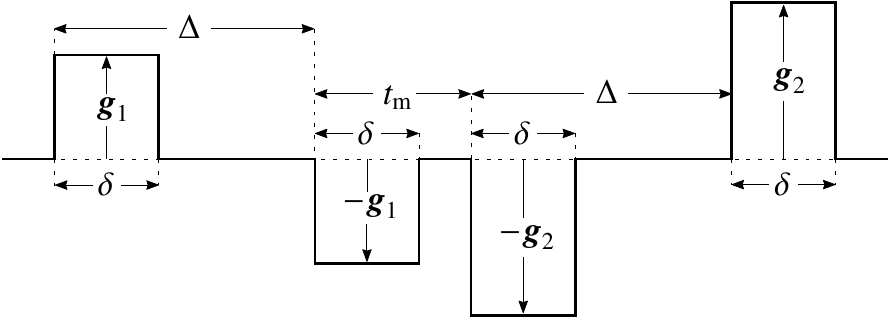}
  \caption{The pulse sequence for double diffusion encoding.} \label{fig:dde}
\end{figure}

As depicted in Fig.\ \ref{fig:dde}, double-diffusion encoding is
achieved by a gradient waveform consisting of two pairs of bipolar
rectangular pulses, each with a given duration $\delta$ and separation
$\Delta$, and magnitudes $\gv_1$ and $-\gv_2$; the minus sign is
customary. The time between the leading edges of the second pulse of the first pair and the
first pulse of the second pair, $\tm$, is called the mixing time. The
calculation of the signal \eqref{eq:signal.conf} therefore entails
very simple integrals, but in a cumbersome piecewise fashion. Upon
significant simplification one finds
\begin{align}
  \Scomp (\gv_1, \gv_2) =
  \exp\!\left(-\gv_1^{\mt} \TS \gv_1 - \gv_2^{\mt} \TS \gv_2
  - 2 \gv_1^{\mt} \TC \gv_2 \right) \,. \label{eq:DDE.conf}
\end{align}
One may refer to the tensors
\begin{subequations}\label{eq:tensors}
\begin{align}
  \TS &= \DD \CO^{-3} \left[ (1-\ee^{-\CO\Delta})(1-\ee^{-\CO\delta})^2
    \ee^{\CO\delta} - (1-\ee^{-2\CO\delta})\ee^{\CO\delta}
    + 2\CO\delta \right] \,, \\
  2\TC &= \DD \CO^{-3} \ee^{-\CO(\tm - \delta)}
    (1-\ee^{-\CO\delta})^2 (1-\ee^{-\CO\Delta})^2 \,,
\end{align}
\end{subequations}
respectively, as the \emph{self-coupling} and \emph{cross-coupling}
tensors between encoding blocks.\footnote{$\TS$ is the same tensor
  that appears in the single-encoding signal $\Scomp = \ee^{-\gv^{\mt}
    \TS \gv}$ \cite{Yolcu16}. In the free diffusion limit ($\CO \to
  0$), it is easily shown that $\TS \to \DD \delta^2 (\Delta -
  \delta/3)$, which recovers the free diffusion signal
  \cite{StejskalTanner65, Stejskal65}. \label{fn:free.self} } For the
orientational average below, we consider $|\gv_1|=|\gv_2|=\gr$, with a
fixed angle $\psi$ between them.

As the free diffusion limit is approached, one can see that the
cross-coupling of encoding blocks vanishes, as $\TC \sim (\DD/2) \CO
\delta^2 \Delta^2 (1 + \CO \delta -\CO \tm)$, due to displacements in
separate time intervals being uncorrelated in pure Brownian motion
\cite{RiskenBook}. The dependence on the mixing time $\tm$ does not
enter until second order in confinement, in line with approximate
calculations done for a spherical wall \cite{OzarslanJCP08}.

\section{Double diffusion encoding: powder average}

The orientational (powder) average of the compartment signal
\eqref{eq:DDE.conf} is performed as follows: The vectors $\gv_1$ and
$\gv_2$ form a plane with unit normal vector
\begin{align}
  \nhat = \frac{\gv_1 \cross \gv_2}{\abs{\gv_1 \cross \gv_2}} \,.
\end{align}
One constructs the average by integrating over all orientations of
(the plane normal to) $\nhat$, and for each of these integrate over
all orientations of the pair $\{\gv_1, \gv_2\}$ within the plane, with
their relative angle $\psi$ fixed. The procedure hence described can
be written as\footnote{We suppress the obvious limits of these
  integrals.}
\begin{align} \label{eq:Sbar.nhat}
  \Sbar = \int \frac{\dd\nhat}{4\pi} \int \frac{\dd\beta}{2\pi}
  \,\Scomp \big(\gv_{\nhat} (\beta), \gv_{\nhat} (\beta+\psi)\big) \,.
\end{align}
Here, $\gv_{\nhat}(\beta)$ is a vector of magnitude $\gr$ in the plane
normal to $\nhat$, whose orientation is parameterized by the
(in-plane) azimuthal angle $\beta$, according to which we can identify
$\gv_1 \to \gv_{\nhat}(\beta)$ and $\gv_2 \to \gv_{\nhat}
(\beta+\psi)$.

We first take the in-plane integral
\begin{align}
  \Sint(\nhat) \stackrel{\rm def}{=} \int \frac{\dd\beta}{2\pi}
  \Scomp \big(\gv_{\nhat} (\beta), \gv_{\nhat} (\beta+\psi)\big)
  \stackrel{\rm def}{=} \int \frac{\dd\beta}{2\pi}
  \ee^{-\sigma_{\nhat}(\beta)} \,, \label{eq:Sint}
\end{align}
which serves as the definitions for the intermediate quantities
$\Sint(\nhat)$ and $\sigma_{\nhat}(\beta)$. According to
\Eqref{eq:DDE.conf}, the latter is given by
\begin{align}
  \sigma_{\nhat}(\beta) = \gv_{\nhat}^{\mt}(\beta) \TS \gv_{\nhat}(\beta)
    + \gv_{\nhat}^{\mt}(\beta+\psi) \TS \gv_{\nhat}(\beta+\psi)
  + 2 \gv_{\nhat}^{\mt}(\beta) \TC \gv_{\nhat}(\beta+\psi) \,.
\end{align}
To anchor the angle coordinate $\beta$, we define the in-plane
cartesian coordinates ${u},{v}$ such that $\uhat \parallel
\gv_{\nhat}(0)$ and $\vhat = \nhat \cross \uhat$, yielding
\begin{align}
  \gv_{\nhat}(\beta) = \uhat \gr \cos\beta + \vhat \gr \sin\beta \,.
\end{align}
After an exercise in trigonometric simplification, and denoting
$\uhat^{\mt} \TS \vhat = \ts_{uv}$ \etc., one finds
\begin{align}
  \sigma_{\nhat} (\beta) = \varsigma_{\nhat} + \rho_{\nhat} \cos2\beta
  -\varrho_{\nhat} \sin2\beta \,,
\end{align}
with
\begin{subequations}
\begin{align}
  \frac {\varsigma_{\nhat}}{\gr^2} &=
  \big(\ts_{uu} + \ts_{vv}\big) + \big(\tc_{uu}+\tc_{vv}\big) \cos\psi \,, \label{eq:varsigma} \\
  \frac {\rho_{\nhat}}{\gr^2} &= \big(\ts_{uu}-\ts_{vv}\big) \frac {1+\cos2\psi}2
  + \ts_{uv} \sin2\psi + \big(\tc_{uu}-\tc_{vv}\big) \cos\psi + 2\tc_{uv} \sin\psi \,,\\
  \frac {\varrho_{\nhat}}{\gr^2} &= \big(\ts_{uu}-\ts_{vv}\big) \frac {\sin2\psi}2
  - \ts_{uv} \big(1+\cos2\psi\big) + \big(\tc_{uu}-\tc_{vv}\big) \sin\psi - 2\tc_{uv} \cos\psi \,.
\end{align}
\end{subequations}
These depend on $\nhat$ through $\uhat$ and $\vhat$, of course.
Hence via \Eqref{eq:Sint},
\begin{align}
  \Sint (\nhat) &= \ee^{-\varsigma_{\nhat}} \int \frac{\dd\beta}{2\pi}
  \ee^{-\rho_{\nhat} \cos2\beta +\varrho_{\nhat}\sin2\beta} \nonumber \\
  &= \ee^{-\varsigma_{\nhat}}
  I_0 \!\left(\sqrt{\rho_{\nhat}^2+\varrho_{\nhat}^2}\right) \,, \label{eq:Sint.eval}
\end{align}
where an integral representation of the modified Bessel function of
order 0 was recognized \cite{DLMF}. The argument of the square root
can be found after a semi-tedious calculation as
\begin{align}
  \rho_{\nhat}^2 + \varrho_{\nhat}^2 = \gr^4 \left[ \big(\ts_{uu}-\ts_{vv}\big) \cos\psi
    + \big(\tc_{uu} - \tc_{vv}\big) \right]^2
  +4 \gr^4 \big(\ts_{uv} \cos\psi + \tc_{uv}\big)^2 \,, \label{eq:argsqrt}
\end{align}
and we rewrite the powder-averaged double-encoding signal
\eqref{eq:Sbar.nhat} via \Eqref{eq:Sint} and \eqref{eq:Sint.eval} as
\begin{align}
  \Sbar = \int \frac{\dd\nhat}{4\pi}\, \ee^{-\varsigma_{\nhat}} I_0
  \!\left(\sqrt{\rho_{\nhat}^2+\varrho_{\nhat}^2}\right)
  \,. \label{eq:Sbar.int}
\end{align}

For the actual evaluation of the integral, explicit expressions in
terms of the polar and azimuthal angles $(\theta,\phi)$ of $\nhat$
need to be substituted,\footnote{$\nhat = (\sin\theta \cos\phi,
  \sin\theta \sin\phi, \cos\theta)$, $\uhat=(\cos\theta \cos\phi,
  \cos\theta \sin\phi, -\sin\theta)$, and $\vhat = (-\sin\phi,
  \cos\phi,0)$.} which are steps we omit here. Eyeballing how entries
of $\TS$ and $\TC$ appear in \Eqref{eq:Sbar.int} via
\Eqref{eq:varsigma} and \eqref{eq:argsqrt}, one notes that it is
useful to define the intermediate tensors
\begin{subequations}\label{eq:mixedtensors}
\begin{align} 
  \MM &= \gr^2 \big(\TS \cos \psi + \TC\big) \\
  \NN &= \gr^2 \big(\TS + \TC \cos \psi\big) \,.
\end{align}
\end{subequations}
Referring to their eigenvalues, in any preferred order (but the same
for both), as $\mm_{i}$ and $\nn_{i}$, and using the following
shorthand
\begin{subequations}  \label{eq:shorthand}
\begin{align}
  \mmm_{ij} &= \mm_{i} - \mm_{j} \,,\\
  \nnn_{ij} &= \nn_{i} - \nn_{j} \,,\\
  \nnp_{ij} &= \nn_{i} + \nn_{j} \,,
\end{align}
\end{subequations}
the powder-averaged signal \eqref{eq:Sbar.int} attains the form
\begin{multline}
  \Sbar = \ee^{-\nnp_{12}} {\int} \frac {\dd\!\cos\theta\,\dd\phi} {4\pi}
  \ee^{-\left(\nnn_{31} + \nnn_{12} \sin^2\!\phi \right) \sin^2\!\theta}
  \label{eq:Sbar.int.exp}  \\
  \mathop\times\, I_0\Big( \scalebox{.75}[.69]{$\sqrt{
      \left(\mmm_{12} \right)^2
      + 2 \mmm_{12} \left[\mmm_{31}
        +(\mmm_{13}+\mmm_{23}) \sin^2\!\phi \right]\sin^2\!\theta
      + \left( \mmm_{13} + \mmm_{21} \sin^2\!\phi \right)^2 \sin^4\!\theta
    }$}\,\Big) \,,
\end{multline}
upon substantial algebraic manipulation.

Ref.\ \cite{Herberthson19} found series expansions for this form in a
different context. We can use its results. Namely,
\begin{subequations} \label{eq:bigresult}
\begin{align}
  \Sbar &= \ee^{-\nnp_{12}} \sum_{k=0}^{\infty} \sum_{m=0}^k {q}_{mk} {Y}_{mk} \,,
  \label{eq:Magnus} \qqtext{where} \\
  {Y}_{mk} &= \sum_{n=0}^\infty \sum_{l=0}^n \frac {\sqrt{\pi} (-1)^n} {l! (n-l)!}
  \frac {\big(\nnn_{31}\big)^{n-l} \big(\nnn_{12}\big)^l } {2^{2(m+l)+1}} \binom {2m+2l} {m+l}
  \frac {(n+k)!} {\left(n+k+\frac12\right)!} \,, \label{eq:Ymk} \qqtext{and}\\
  {q}_{mk} &= \frac {\big(\mmm_{31}\big)^{\!k{-}m}}{(k{-}m)!}
  \sum_{j=0}^{k/2} \frac {\text{$\big(\mmm_{12}\big)^j I_{k{-}j} \big(\mmm_{12}\big)$}}
      {\text{\small $2^j \! j!\big(\mmm_{13}{+}\mmm_{23}\big)^{\!2j{-}m}$}}
  \,{}_2\!\tilde{F}_1 \!\!\left( \text{$m{-}k, {-}2j;m{+}1{-}2j;$} \tfrac {\mmm_{13}{+}\mmm_{23}}{\mmm_{12}}\!\right) \,, \label{eq:qmk}
\end{align}
\end{subequations}
with the following three alternatives for \Eqref{eq:Ymk}:
\begin{subequations} \label{eq:Y.alt}
\begin{align}
  \label{eq:Y1} {Y}^{(1)}_{mk} &= \frac {\sqrt{\pi}} {2^{2m+1}} \binom{2m}m
  \sum_{n=0}^\infty \frac {\big(\nnn_{13}\big)^n (k+n)!}
      {n! \big(k+n+\frac12\big)!} \,
      {}_2F_1 \!\left(m+\tfrac12, -n; m+1; \tfrac {\nnn_{21}}{\nnn_{31}} \right) \,, \\
  \label{eq:Y2} {Y}^{(2)}_{mk} &= \frac {\sqrt{\pi}} {2^{2m+1}}
    \sum_{n=0}^\infty \frac {\big(\nnn_{21}\big)^n (k+n)! \binom{2n+2m}{n+m}}
     {2^{2n} n! \big(k+n+\frac12\big)!} \,
     {}_1F_1 \!\left(k+n+1; k+n+\tfrac32;\nnn_{13} \right)\,, \\
  \label{eq:Y3} {Y}^{(3)}_{mk} &= \frac {\sqrt{\pi}} {2^{2m+1}} \binom{2m}m
  \sum_{n=0}^\infty \frac {\big(\nnn_{13}\big)^{\!n} (k{+}n)!}
   {n! \big(k{+}n{+}\frac12\big)!} \,
   \!{}_2F_2 \!\!\left(m{+}\tfrac12, k{+}n{+}1; m{+}1, k{+}n{+}\tfrac32; \nnn_{21} \right)\,.
\end{align}
\end{subequations}
Here, ${}_iF_{\!j} (\ldots)$ are the confluent hypergeometric
functions \cite{Arfkenbook}, tilde denoting regularization. Note that
these expressions apply to the most general case, in which the
confinement tensor (and thus the tensors that are functions of it)
have three \emph{distinct} eigenvalues. Special cases associated with
coinciding eigenvalues are discussed in the appropriate occasion
later.

Which alternative among \Eqref{eq:Y.alt} yields better convergence is
a matter of how the eigenvalues $\nn_i$ are chosen to be ordered, by
way of the sizes and signs of $\nnn_{ij}$. As a general guideline it
would be wise to order the eigenvalues so as to avoid a sequence that
alternates in sign, and with a large expansion parameter. Take
\Eqref{eq:Y2} for instance. Given that
${}_1F_1(k{+}l{+}1;k{+}l{+}\frac32;\nnn_{13}) > 0$ and increasing for
all $\nnn_{13}$, it would be beneficial to make $\nnn_{13}$ negative (and
large if possible), while keeping $\nnn_{21}$ positive (and small if
possible). For a given set of eigenvalues $\nn_i$, ordering them in
the fashion $\nn_1 < \nn_2 <\nn_3$ would be along this guideline,
whereas the ordering $\nn_2 < \nn_3 < \nn_1$ would result in a
sequence with larger terms and alternating sign.

  
Note however that while $\TS$ is a monotonic (decreasing) function of
$\CO$, $\TC$ is not; see \Eqref{eq:tensors}. Through
\Eqref{eq:mixedtensors} this means that their mixtures $\MM$ and $\NN$
are not necessarily monotonic in the confinement $\CO$. Hence what
ordering of the confinement eigenvalues $\co_i$ achieves what ordering
in the eigenvalues $\mm_i$ and $\nn_i$ is a question which has an
answer only on a case-by-case basis. Furthermore, the ordering of
$\co_i$ that yields a desirable ordering of $\nn_i$ for a particular
one of \Eqref{eq:Y.alt} may not produce an ordering of $\mm_i$ as
desirable for the convergence of ${q}_{mk}$ in \Eqref{eq:qmk}.

\subsection{Axisymmetric confinement}

We refer to the condition when two of the eigenvalues $\co_i$ of the
confinement tensor coincide as \emph{axisymmetric}
confinement. Under this condition, the series expansions above undergo
simplifications.

The most drastic simplification occurs when $\co_1=\co_2$. That is,
given that two eigenvalues coincide, assigning first and second place
to them is the wisest choice as far as the evaluation of the series
expansions \eqref{eq:bigresult} is concerned.

First, the coefficient ${q}_{mk}$ simplifies as\footnote{For this, it
  needs to be noted \cite[supplementary information]{Herberthson19}
  that the way ${q}_{mk}$ arises in the calculation---before ever
  arriving at \Eqref{eq:qmk}---is that it is the coefficient in the
  (double) series expansion of the Bessel function in
  \Eqref{eq:Sbar.int.exp}: \[I_0 (\ldots) =
  \sum_{k=0}^{\infty}\sum_{m=0}^{k} {q}_{mk} \sin^{2m}\!\phi\,
  \sin^{2k}\!\theta \,.\]Upon all coefficients in the argument except for
  $\mmm_{13}$ vanishing due to axisymmetry, one has \[I_0 \big(\mmm_{13}
  \sin^2\!\theta\big) = \sum_{k=0}^{\infty}
  \frac{\smash[t]{\big(\mmm_{13}\big)}^{2k}} {(k!)^2 2^{2k}}
  \sin^{2k}\!\theta\,,\]which, comparing to the previous (double)
  expansion, implies \Eqref{eq:q.axy}.}
\begin{align} \label{eq:q.axy}
  {q}_{mk} &= \begin{cases} \frac {\delta_{m0} (\mmm_{13})^{k}}
    {(\frac k2)! (\frac k2)! 2^k} \;,
      &\text{even}\; k \\ 0 \;, & \text{odd} \; k \end{cases}\;.
\end{align}
The coefficient ${Y}_{mk}$, on the other hand, loses the interior
summation in \Eqref{eq:Ymk} due to $\mmm_{12}$ vanishing. The
remaining summation can be identified according to the definition of
hypergeometric functions \cite{Arfkenbook} as
\begin{align}
  {Y}_{mk} = \frac {\sqrt{\pi}} {2^{m+1}} \binom{2m}{m}
  \frac{k!}{\big(k+\frac12\big)!}  \, {}_1F_1 \!\left(
  k+1; k+\tfrac32; \nnn_{13} \right) \,, \label{eq:Y.axy}
\end{align}
which yields via \Eqref{eq:q.axy} and \Eqref{eq:Magnus}
\begin{align}
    \Sbar_{\rm axy} &= \sqrt{\pi} \ee^{-\nnp_{12}} \sum_{n=0}^\infty
  \binom{2n}n \frac{\smash[t]{\big(\mmm_{13}\big)}^{2n}} {2^{2n+1}}
  \frac {{}_1F_1 \!\left(2n+1; 2n+\tfrac32;
    \nnn_{13}\right)} {\big(2n+\frac12\big)!} \,.\label{eq:Sbar.axy}
\end{align}
This expansion is not so sensitive to the ordering of eigenvalues
number 1 and 3, as the expansion parameter ($\mmm_{13}$) is squared,
and the hypergeometric function ${}_1F_1(2n{+}1;2n{+}3/2; \nnn_{13})
>0$ for all arguments.


\begin{figure}[t!]
  \centering\includegraphics[scale=1.1]{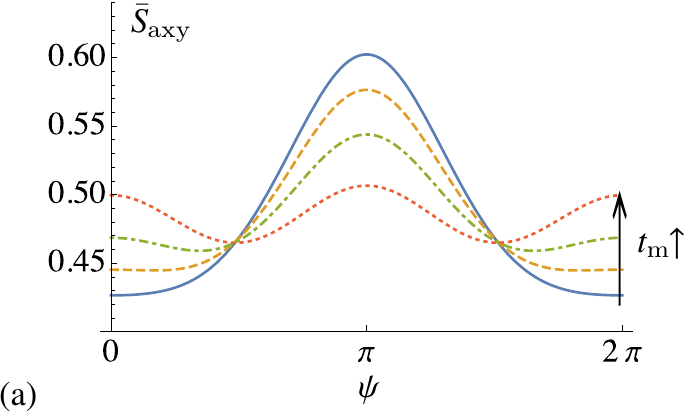} \hspace{.5cm}
  \includegraphics[scale=1.1]{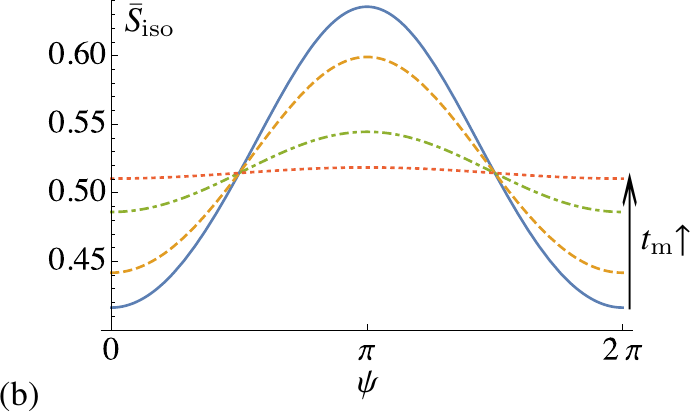} \\[10pt]
   \includegraphics[scale=1.1]{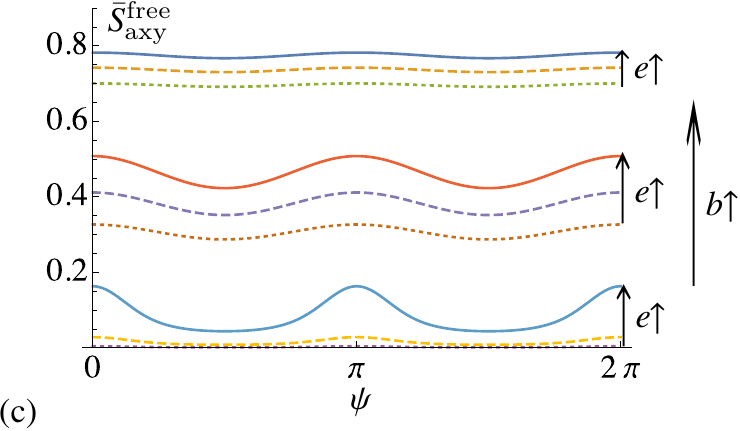}
  \caption{Dependence of powder averaged signal \eqref{eq:Sbar.axy} on
    the angle $\psi$ between double-encoding gradient directions. (a)
    Prolate confinement $\CO$, with eigenvalues $\{1,1,0.1\}$ in
    arbitrary units. $\Delta=10$ and $\delta=1$, in units of
    $\co_1^{-1}$. The mixing time $\tm$ was set to the following
    multiples $\{1, 2, 6, 30\}$ of $\delta$. The gradient strength is
    set such that $D_0 \gr^2 /\co_1^3=0.16$, with $D_0$ being the unit
    of diffusivity. (b) Isotropic confinement $\CO$, with eigenvalues
    $\{1,1,1\}$ in arbitrary units. $\Delta=3$ and $\delta=2/3$, in
    units of $\co_1^{-1}$. The mixing time $\tm$ was set to the
    following multiples $\{1, 3/2, 3, 6\}$ of $\delta$. The gradient
    strength is set such that $D_0 \gr^2 /\co_1^3=1$. (c) Free
    diffusion with varying anisotropy (prolate) for comparison, with
    diffusivity $\DD$ eigenvalues $\{D_0, {e} D_0, {e}
    D_0\}$. Across the three groups of plots, diffusion weighting
    ${b}=\gr^2 \delta^2 (\Delta-\delta/3)$ takes the values $\{0.4,
    1.4, 10\}$ in units of $D_0^{-1}$, whereas each group of plots
    have the eccentricity parameter set to $\{0.01, 0.1,
    0.2\}$.} \label{fig:tm}
\end{figure}

We depict in Fig.\ \ref{fig:tm} the evaluation of the powder averaged
signal \eqref{eq:Sbar.axy} for axisymmetric confinement for
representative values of encoding parameters. In
Fig.\ \ref{fig:tm}(a), the confinement is anisotropic (prolate),
whereas in Fig.\ \ref{fig:tm}(b), it is (nearly) isotropic. The
bell-shaped dependence on the relative angle $\psi$ between the
gradient directions is seen in both cases, which is a sign that
diffusion is not free \cite{OzarslanJCP08}. This dependence is due
mainly to the exponential prefactor in \Eqref{eq:Sbar.axy} that has
nothing to do with the difference between the confinement eigenvalues
(anisotropy). When the mixing time is increased, the bell-shaped
modulation, indicating confinement regardless of anisotropy, stops
overwhelming the relatively smaller influence of the rest of the
expression \eqref{eq:Sbar.axy}: see Fig.\ \ref{fig:tm}(a) where the
confinement is anisotropic. In an isotropic confinement, on the other
hand, angular modulation simply disappears when the mixing time is
increased (Fig.\ \ref{fig:tm}b), illustrating that the angular
modulation that survives the increase in mixing time is due only to
compartmental anisotropy, and not due to the fact of confinement (or
to ensemble anisotropy, which was already eliminated by powder
averaging).




\subsection{Insights from two dimensions}

The signal expressions for double encoding are a bit unwieldy to get a
conceptual handle on. However, some insight can be gleaned from
considering the orientational averaging in two dimensions.

Obviously, in the spirit of \Eqref{eq:Sbar.nhat}, the 2D
orientationally averaged signal can be written as
\begin{align} 
  \Sbar_{2\rm D} = \int \frac{\dd\beta}{2\pi}
  \,\Scomp (\gv (\beta), \gv (\beta+\psi)) \,.
\end{align}
That is, there is no normal vector to integrate over; everything takes
place in an $(x,y)$ plane. The same steps \Eqref{eq:Sint} through
\Eqref{eq:Sint.eval} apply, and one has
\begin{align}
  \Sbar_{2\rm D} &= \ee^{-\gr^2 \Tr \big(\TS+\TC \cos\psi \big)}
    I_0 \big(\gr^2 (\ts_1 - \ts_2)\cos\psi + \gr^2 (\tc_1-\tc_2) \big) \,.
\end{align}
The dependence on the relative angle $\psi$ of the gradient directions
occurs both in the exponential attenuation factor and in the Bessel
function. The angular dependence in the Bessel function has to do with
anisotropy ($\ts_1-\ts_2$) which the exponential factor is insensitive
to due to the trace. In the exponent, on the other hand, the angular
dependence is controlled by $\Tr \TC$, whose presence is due to
confinement (since $\TC \to 0$ in the free diffusion limit). For large
mixing times ($\tm \co_i \gg 1$) the latter drops out and the angular
modulation due to anisotropy is liberated. However, for smaller mixing
time, it turns out that the angular dependence of the exponential
factor dominates, which is due to confinement, suppressing the
signature of anisotropy.

An interesting double-encoding scheme is its ``symmetrized'' version \cite{Paulsen15}, which fixes $\psi=\pi/2$, but varies the magnitudes
$\gr_1 = \gr \cos \alpha$ and $\gr_2 = \gr \sin \alpha$ as a function
of some parameter $\alpha$. In that scenario, one can easily calculate
that
\begin{align}
  \Sbar_{2\rm D}^{\rm sym} &= \ee^{- \frac{\gr^2}{2} \Tr (\TS)}
  I_0 \!\left( \frac{g^2}{2\sqrt{2}} \sqrt{
    \big( \Delta\tss^2 + \Delta\tcc^2 \big) +
    \big( \Delta\tss^2 - \Delta\tcc^2 \big) \cos4\alpha}\right) \,,
\end{align}
showing that the modulation due to confinement in the exponential factor drops out, and one of
pure anisotropy remains (with $\Delta \tss$ denoting $\ts_1-\ts_2$
\etc.). The result is a `cleaner' version of the signal modulation wherein the confinement anisotropy is the only source of angular modulation characterized by the $\cos 4\alpha$ dependence. It should be remembered though that the confinement model is purely Gaussian, as such does not account for compartmental kurtosis. When truly restricted diffusion is considered, the compartment anisotropy and compartmental kurtosis both yield the same type of angular dependence compromising the interpretation of such angular dependence except when the compartments are isotropic \cite{Paulsen15}.

\section{Single diffusion encoding}

The powder-averaged single-encoding signal can be obtained by way of
``hacking'' its double-encoded counterpart. One notes that $\Scomp =
\ee^{-\gv^{\mt} \TS \gv}$ is the single-encoding signal at the
compartment level \cite{Yolcu16}. Then the features of the form
\eqref{eq:DDE.conf} to get rid of are $(i)$ the presence of a second
vector $\gv_2$ unequal to the first, $(ii)$ the presence of
cross-coupling ($\TC$), and $(iii)$ the double occurrence of
self-coupling. The first is hacked away by setting $\psi=0$. For the
second, one simply sets $\TC=0$.\footnote{Physically, this can be
  imagined as the limit $\tm \to \infty$. However, physics is not
  necessary. We simply have a set of expressions,
  \eg.\ \Eqref{eq:bigresult}, containing $\TS$ and $\TC$ via
  \Eqref{eq:mixedtensors} via \Eqref{eq:shorthand}, and we want to
  remove instances of $\TC$.} The result is $\Scomp = \ee^{-2
  \gv^{\mt} \TS \gv}$, suffering from the third problem above, which
is fixed by replacing $\gr \to \gr/\sqrt{2}$. All this yields via
\Eqref{eq:mixedtensors}
\begin{align} \label{eq:doublesingle}
  \MM = \tfrac12 \gr^2 \TS = \NN \,,
\end{align}
which is the substitution that converts the powder-average expressions
for double-encoding into those of single-encoding.

With this condition applied, the $\nnn_{ij}$'s appearing in
\Eqref{eq:bigresult} turn into $\mmm_{ij}$'s. However this alone does
not produce drastic simplifications such as reducing
summations. Rather than using \Eqref{eq:bigresult}, in fact, it is
better to note that other results in Ref.\ \cite{Herberthson19} are
quite suitable in the case of single-encoding. Since the compartment
signal has the form $\Scomp = \ee^{-\Tr \left( \TS \gv \gv^{\mt}
  \right)}$, with the matrix $\gv \gv^{\mt}$ axisymmetric by dint of
being rank-1, the results of Ref.\ \cite{Herberthson19} for
axisymmetric diffusion or measurement tensor can be applied. Written
in terms of the parameters of the present discussion, the relevant
formulas of Ref.\ \cite{Herberthson19} indicate the following
alternative expressions
\begin{subequations} \label{eq:powder.axy}
\begin{align}
  \Sbar^{(1)} &= \frac {\sqrt{\pi}}2 \ee^{-\gr^2 \ts_3}
    \sum_{n=0}^\infty \frac {\gr^{2n} \big(\ts_3-\ts_1\big)^n} {\big(n+\frac12\big)!}
        \,{}_2F_1 \Big(\tfrac12,-n;1;\tfrac {\ts_1-\ts_2}{\ts_1-\ts_3}\Big) \,, \\
  \Sbar^{(2)} &= \ee^{-\gr^2 \ts_3} \sum_{n=0}^\infty
  \frac {\gr^{2n} \big(\ts_1-\ts_2\big)^n}{n!(2n+1)!}
        \,{}_1\!F_1\big(n+1;n+\tfrac32;\gr^2(\ts_3-\ts_1)\big) \,, \\
  \Sbar^{(3)} &= \frac {\sqrt{\pi}}2 \ee^{-\gr^2 \ts_3}
        \sum_{n=0}^\infty \frac {\gr^{2n}\big(\ts_3-\ts_1\big)^n} {\big(n+\frac12\big)!}
          \, {}_2F_2\big(\tfrac12,n+1;1,n+\tfrac32;\gr^2(\ts_1-\ts_2)\big) \,,\\
  \Sbar^{(4)} &= \frac {\sqrt{\pi}}2 \ee^{\!-\gr^2 \ts_3}
     \!  \sum_{n=0}^\infty \frac {\binom{2n}{n}\gr^{4n}\big(\ts_1-\ts_2\big)^{\!2n}}
          {16^n \big(2n{+}\frac12\big)!}  
      {}_1\!F_1 \!\big(2n{+}1;2n{+}\tfrac32;\tfrac{\gr^2}2 (2\ts_3{-}\ts_2{-}\ts_1)\big) ,
\end{align}
\end{subequations}
for fully anisotropic $\CO$ (hence $\TS$); the axisymmetric case is
taken up below.  These summations are subject to the same guidelines
that followed \Eqref{eq:Y.alt} except that the caveats about
conflicting eigenvalue ordering do not apply here. The single-encoding
expressions here only involve the self-coupling tensor $\TS$ which is
a monotonically decreasing function of $\CO$; see
\Eqref{eq:tensors}. Therefore the ordering of the eigenvalues $\ts_i$
and $\co_i$ are certain to be in exactly the opposite sense of each
other.

\subsection{Axisymmetry and the power-laws for confined diffusion}

The first alternative in \Eqref{eq:powder.axy} needs special care when
$\ts_1=\ts_3$ and the argument of the hypergeometric function ${}_2F_1
(\ldots)$ diverges. Invoking a property of the hypergeometric
function,\footnote{\[\lim_{x\to 0} x^n {}_2F_1 \!\left( m{+}\tfrac12,
  {-}n;m{+}1;\tfrac yx\right) = (-y)^n \frac
  {\big(m+n-\frac12\big)!\,m!}{\big(m-\frac12\big)!\, (m+n)!} \,.\]}
and renaming $\ts_1=\ts_{\perp}$ and $\ts_2=\ts_{\parallel}$, one finds\footnote{
  Alternatively to using the results of Ref.\ \cite{Herberthson19}
  here, one may go back to the integral expression
  \eqref{eq:Sbar.int.exp}. First, we note that axisymmetry, with the
  choice $\co_1=\co_2$, makes the integrand independent of $\phi$,
  yielding
\begin{align*}
  \Sbar &= \tfrac12 \ee^{-\nnp_{12}} {\int} \dd\!\cos\theta\,
        \ee^{\nnn_{13}\sin^2\!\theta}
        I_0\!\left(\mmm_{13} \sin^2\!\theta \right) \,.
\end{align*}
We are not aware of a closed-form evaluation of this
integral. However, its special case \eqref{eq:doublesingle} relevant
here, making the arguments of the exponential and Bessel function
match, has the result \cite[6.625-4]{Gradshteynbook} 
\begin{align*}
  \Sbar &= \frac {\sqrt{\pi}}2 \ee^{-2\mm_1}
  \frac {\erf\!\Big(\sqrt{2\mmm_{31}}\Big)} {\sqrt{2\mmm_{31}}}
  = \frac {\sqrt{\pi}}2 \ee^{-\gr^2 \ts_1}
  \frac {\erf\!\Big(\gr\sqrt{\ts_3 - \ts_1}\Big)}
        {\gr\sqrt{\ts_3 - \ts_1}} \,.
\end{align*}}
\begin{align} \label{eq:single.axy}
  \Sbar_{\rm axy} &= \frac {\sqrt{\pi}}2 \ee^{-\gr^2 \ts_{\perp}(\delta,\Delta)}
  \frac {\erf\!\Big( \gr \sqrt{\phantom{\ts_1}\mkern-18mu
      \smash[b]{\ts_{\parallel}(\delta,\Delta) - \ts_{\perp}(\delta,\Delta)}} \Big)}
        {\gr \sqrt{\phantom{\ts_1}\mkern-18mu
      \smash[b]{\ts_{\parallel}(\delta,\Delta) - \ts_{\perp}(\delta,\Delta)}}} \,.
\end{align}
Here, we have also explicitly denoted the dependence on the encoding
protocol's timing parameters as per \Eqref{eq:tensors}.
Note that in the free diffusion limit ($\CO \to 0$, see footnote
\ref{fn:free.self}), the known single-encoding powder-average signal
\cite{Kroenke04, Yablonskiy02, Anderson05}
\begin{align}
  \Sbar_{\rm axy} &= \frac {\sqrt{\pi}}2
  \ee^{-\qr^2 (\Delta-\frac\delta3) {D}_\perp}
  \frac {\erf\!\Big( \qr \sqrt{\big(\Delta-\smash{\frac\delta3}\big)
      \big( {D}_{\parallel} - {D}_{\perp} \big)} \Big)}
        {\qr \sqrt{\big(\Delta-\smash{\frac\delta3}\big)
      \big( {D}_{\parallel} - {D}_{\perp} \big)} } \,,
\end{align}
with $\qr = \gr \delta$, is recovered.

Two special limiting cases follow.

\subsubsection{Stick: $\Sbar \propto g^{-1}$ scaling }
When particles have negligible latitude to move in the transverse
direction, the orientationally-averaged signal \eqref{eq:single.axy} assumes a
special form. In terms of the confinement model, this corresponds to
$\co_\perp \to \infty$, which implies $\ts_\perp \to 0$ via
\Eqref{eq:tensors}. The signal \eqref{eq:single.axy} then becomes
\begin{align}
  \Sbar_{\rm stick} &= \frac {\sqrt{\pi}}2
  \frac {\erf\!\Big(\gr \sqrt{\phantom{\ts_1}\mkern-18mu\smash[b]{\ts_\parallel(\delta,\Delta)}}\Big)}
        {\gr \sqrt{\phantom{\ts_1}\mkern-18mu\smash[b]{\ts_\parallel(\delta,\Delta)}}}
  \stackrel{\gr\to\infty}{\sim}
  \frac {\sqrt{\pi}}
        {2\gr\sqrt{\phantom{\ts_1}\mkern-18mu\smash[b]{\ts_\parallel(\delta,\Delta)}}}
  \stackrel{\co_{\parallel} \to 0}{\sim}
  \frac {\sqrt{\pi}}
        {2\qr \sqrt{\big(\Delta-\smash{\frac\delta3}\big) {D}_{\parallel}}}
  \,. \label{eq:stick}
\end{align}
The large gradient regime has been important in identifying stick-like compartments via the $\qr^{-1}$ scaling, which has been observed in white-matter areas of the brain and has been interpreted with the assumption of a free one-dimensional diffusion \cite{McKinnon17}, which is adequate for channels of straight long channels of infinitesimal diameter. A notable exception is Ref. \cite{Ozarslan18FiP}, which has incorporated the effects of finite size and curvature. Ref. \cite{Ozarslan18FiP} also pointed out that such a scaling is not the true asymptotic behavior of the signal; the latter is rather dictated by the Debye-Porod law for narrow pulses \cite{Sen95} and an even steeper attenuation is predicted when the pulses are wide. 

The above expression suggests that a similar decay is expected for the gradient magnitude rather than the $q$-value. The crucial difference is in the dependence on the timing parameters $\{\delta,\Delta\}$ of the SDE sequence, see Figure 1. It would thus be interesting, e.g. in white-matter, to investigate whether the dependence on the timing parameters is more like
$1/\sqrt{\Delta-\delta/3 }$ (free difusion along the fiber) or
$1/\sqrt{\phantom{\ts_1}\mkern-18mu\smash[b]{\ts_\parallel(\delta,\Delta)}}$
(confined diffusion along the fiber), which would inform about the diffusion process along the axons.

\subsubsection{Pancake: $\Sbar \propto g^{-2}$ scaling}

For completeness, we consider the opposite case where particles are able to spread in a plane, but not along the normal. Here, $\co_\parallel \to \infty$, implying $\ts_\parallel \to 0$ via \Eqref{eq:tensors}. One then has\footnote{The imaginary error function has the properties $\im
  \erfi (z) = \erf (\im z)$, and $\erfi (z) \sim \ee^{z^2}\!/z\sqrt{\pi}$. }
\begin{align}
  \Sbar_{\rm pancake} &= \frac {\sqrt{\pi}}2 \ee^{-\gr^2 \ts_\perp(\delta,\Delta)}
  \frac {\erfi \Big( \gr \sqrt{\ts_\perp(\delta,\Delta)} \Big)}
        {\gr \sqrt{\ts_\perp(\delta,\Delta)}}
  \stackrel{\gr\to\infty} {\sim}
  \frac 1{2\gr^2\ts_\perp(\delta,\Delta)}
  \stackrel{\co_\perp \to 0}{\sim}
  \frac1{2\qr^2 \big(\Delta-\frac\delta3\big){D}_{\perp}}\,.
\end{align}
Thus, the orientationally-averaged signal attenuates at a faster rate than in the case of sticks. Similarly though, the dependence on the timing parameters are different in free and confined diffusion scenarios.

\section{Discussion}

We have provided, for the first time, explicit expressions for the orientationally-averaged SDE and DDE  MR signal intensity for structures represented by confinement tensors \cite{Yolcu16}. The latter is the effective model of restricted diffusion when pulses are long enough for the diffusing particles to traverse distances larger than the pore size \cite{Ozarslan17FiP}. As such, our findings are relevant for a broad range of porous materials featuring isolated, small pores. 

The counterpart of these results for compartments of free anisotropic diffusion were given in Ref.\ \cite{Herberthson19} for arbitrary encoding waveforms. For the time being, considering arbitrary waveforms for confined compartments seems extremely challenging. However, taking confinement into account is important, since the free diffusion model lacks features (such as the dependence on relative angle in double encoding) that a realistic signal will bear, see Figure 3. Such bell-shaped angular modulation was related to the radius of gyration of the pores \cite{Mitra95} and has since been used to estimate the apparent size of pores \cite{OzarslanJCP08,Shemesh09,OzarslanMRC11,Callaghanbook2,Komlosh13NI} via DDE measurements. The underlying reason  was thought to be a restriction effect \cite{Finsterbusch2011review} that not only  leads to an anisotropy of the diffusion process at a length scale smaller than the pore size \cite{OzarslanBJ08,OzarslanJCP08,OzarslanJCP09,Moutal19tmi}, but also makes the apparent diffusion coefficient depend on the diffusion time \cite{Jespersen12}. Thus, this effect is absent when free diffusion is thought to take place within individual pores, see Figure 3c. Our results indicate that the confinement tensor framework is capable of capturing such angular modulation, which makes it a suitable representation of diffusion within microdomains \cite{Yolcu16,Ozarslan17FiP,LiuOzarslan2019NBMreview}. This is important in applications like $q$-space trajectory imaging \cite{Westin16_QTI} and diffusion tensor distribution imaging \cite{Topgaard19DTDI}, which aim to characterize the structure of subdomains using general gradient waveforms.

Another angular modulation that is apparent in Figure 3 is  w-shaped, which was pointed out by Mitra \cite{Mitra95} for randomly distributed sticks. Later, it was proved that the dominant contribution to such angular modulation had the functional form $\cos 2\psi$ for fully restricted structures and cylinders of finite diameter \cite{Ozarslan_JMR09}. Such modulation that manifests itself at twice the ``angular frequency'' \cite{Mitra95,Ozarslan_JMR09,Lawrenz10} is present even when diffusion within the subdomains is envisioned to be free as long as it is anisotropic, see Figure 3c. Thus, such modulation is truly indicative of the  anisotropy of the subdomains, be it free (Figure 3c), confined (Figure 3a\&b), or truly restricted  \cite{Ozarslan_JMR09,OzarslanJCP09}. For more information on such anisotropy, the reader is referred to the review on this topic by Ianu\c s et al. \cite{Ianus17MAVAreview} in this book series.


From a mathematical point of view, the expressions in Ref.\ \cite{Herberthson19} provided the Laplace transform of a tensor distribution, which includes rotated copies of a given diffusion tensor wherein all orientations are equally likely, thus extending the modeling approach that employs parametric diffusion tensor distributions \cite{Jian07,Leow09,Scherrer16,Shakya17Dagstuhl} to a new type of tensor distribution. Evaluating the signal, in a similar fashion, for confinement rather than diffusion tensors can be regarded as the evaluation of a transform whose kernel is the compartmental signal given in \eqref{eq:DDE.conf} instead of the kernel of the matrix Laplace transform $\ee^{-\mathbf{B}\mathbf{D}}$.

\section{Conclusion}

We have given analytical expressions for the orientationally averaged diffusion MR signal originating from confined anisotropic compartments for two relatively simple encoding schemes. A number of observations related to signal modulation and power-law tails were made for such confined pores. These findings complement and extend the exact expressions for locally free diffusion provided in Ref.\ \cite{Herberthson19} to confined diffusion albeit for SDE and DDE measurements.

\section*{Acknowledgments}

We acknowledge the following sources for funding: Swedish Foundation for Strategic Research AM13-0090, the Swedish Research Council 2016-04482, Linköping University Center for Industrial Information Technology (CENIIT), VINNOVA/ITEA3 17021 IMPACT, and National Institutes of Health P41EB015902 and R01MH074794.

\bibliography{../../sharedbib}

\begin{thebibliography}{10}

\bibitem{StejskalTanner65}
E.~O. Stejskal and J.~E. Tanner, ``Spin diffusion measurements: Spin echoes in
  the presence of a time-dependent field gradient,'' {\em J Chem Phys},
  vol.~42, no.~1, pp.~288--292, 1965.

\bibitem{Cory90b}
D.~G. Cory, A.~N. Garroway, and J.~B. Miller, ``Applications of spin transport
  as a probe of local geometry,'' {\em Polym Preprints}, vol.~31, p.~149, 1990.

\bibitem{ChengCory99}
Y.~Cheng and D.~G. Cory, ``Multiple scattering by {NMR},'' {\em J Am Chem Soc},
  vol.~121, pp.~7935--7936, 1999.

\bibitem{Callaghanbook2}
P.~T. Callaghan, {\em Translational dynamics and magnetic resonance: Principles
  of pulsed gradient spin echo {NMR}}.
\newblock New York: Oxford University Press, 2011.

\bibitem{Finsterbusch2011review}
J.~Finsterbusch, ``Multiple-wave-vector diffusion-weighted {NMR},'' {\em Annual
  Reports on NMR Spectroscopy}, vol.~72, pp.~225--299, 2011.

\bibitem{Shemesh12}
N.~Shemesh, E.~\"Ozarslan, P.~J. Basser, and Y.~Cohen, ``Accurate noninvasive
  measurement of cell size and compartment shape anisotropy in yeast cells
  using double-pulsed field gradient {MR}.,'' {\em NMR Biomed}, vol.~25,
  pp.~236--246, Feb 2012.

\bibitem{Shemesh16nomenclature}
N.~Shemesh, S.~N. Jespersen, D.~C. Alexander, Y.~Cohen, I.~Drobnjak, T.~B.
  Dyrby, J.~Finsterbusch, M.~A. Koch, T.~Kuder, F.~Laun, M.~Lawrenz,
  H.~Lundell, P.~P. Mitra, M.~Nilsson, E.~{\"O}zarslan, D.~Topgaard, and C.-F.
  Westin, ``Conventions and nomenclature for double diffusion encoding {NMR}
  and {MRI},'' {\em Magn Reson Med}, vol.~75, pp.~82--7, Jan 2016.

\bibitem{Novikov19NBMreview}
D.~S. Novikov, E.~Fieremans, S.~N. Jespersen, and V.~G. Kiselev, ``Quantifying
  brain microstructure with diffusion {MRI}: Theory and parameter estimation,''
  {\em NMR Biomed}, vol.~32, p.~e3998, 04 2019.

\bibitem{RiskenBook}
H.~Risken, {\em The Fokker-Planck Equation}.
\newblock Springer-Verlag, 2~ed., 1989.

\bibitem{Mitra95}
P.~P. Mitra, ``Multiple wave-vector extensions of the {NMR}
  pulsed-field-gradient spin-echo diffusion measurement,'' {\em Phys Rev B},
  vol.~51, no.~21, pp.~15074--15078, 1995.

\bibitem{Callaghan02}
P.~T. Callaghan and M.~E. Komlosh, ``Locally anisotropic motion in a
  macroscopically isotropic system: displacement correlations measured using
  double pulsed gradient spin-echo {NMR},'' {\em Magn Reson Chem}, vol.~40,
  pp.~S15--S19, 2002.

\bibitem{OzarslanJCP08}
E.~\"Ozarslan and P.~J. Basser, ``Microscopic anisotropy revealed by {NMR}
  double pulsed field gradient experiments with arbitrary timing parameters,''
  {\em J Chem Phys}, vol.~128, no.~15, p.~154511, 2008.

\bibitem{Uhlenbeck30}
G.~E. Uhlenbeck and L.~S. Ornstein, ``On the theory of the {B}rownian motion,''
  {\em Phys Rev}, vol.~36, pp.~823--841, 1930.

\bibitem{Callaghan80}
P.~T. Callaghan and D.~N. Pinder, ``Dynamics of entangled polystyrene solutions
  studied by pulsed field gradient nuclear magnetic resonance,'' {\em
  Macromolecules}, vol.~13, pp.~1085--1092, 1980.

\bibitem{LeDoussal92}
P.~{Le Doussal} and P.~N. Sen, ``Decay of nuclear magnetization by diffusion in
  a parabolic magnetic field: An exactly solvable model,'' {\em Phys Rev B},
  vol.~46, no.~6, pp.~3465--3485, 1992.

\bibitem{MitraHalperin95}
P.~P. Mitra and B.~I. Halperin, ``Effects of finite gradient-pulse widths in
  pulsed-field-gradient diffusion measurements,'' {\em J Magn Reson A},
  vol.~113, pp.~94--101, 1995.

\bibitem{Yolcu16}
C.~Yolcu, M.~Memi\c{c}, K.~\c{S}im\c{s}ek, C.~F. Westin, and E.~\"Ozarslan,
  ``{NMR} signal for particles diffusing under potentials: From path integrals
  and numerical methods to a model of diffusion anisotropy,'' {\em Phys Rev E},
  vol.~93, p.~052602, 2016.

\bibitem{Ozarslan17FiP}
E.~\"{O}zarslan, C.~Yolcu, M.~Herberthson, C.-F. Westin, and H.~Knutsson,
  ``Effective potential for magnetic resonance measurements of restricted
  diffusion,'' {\em Front Phys}, vol.~5, p.~68, 2017.

\bibitem{Stejskal65}
E.~O. Stejskal, ``Use of spin echoes in a pulsed magnetic-field gradient to
  study anisotropic, restricted diffusion and flow,'' {\em J Chem Phys},
  vol.~43, no.~10, pp.~3597--3603, 1965.

\bibitem{DLMF}
``{\it NIST Digital Library of Mathematical Functions}.''
  http://dlmf.nist.gov/, Release 1.0.24 of 2019-09-15.
\newblock F.~W.~J. Olver, A.~B. {Olde Daalhuis}, D.~W. Lozier, B.~I. Schneider,
  R.~F. Boisvert, C.~W. Clark, B.~R. Miller, B.~V. Saunders, H.~S. Cohl, and
  M.~A. McClain, eds.

\bibitem{Herberthson19}
M.~Herberthson, C.~Yolcu, H.~Knutsson, C.-F. Westin, and E.~{\"O}zarslan,
  ``Orientationally-averaged diffusion-attenuated magnetic resonance signal for
  locally-anisotropic diffusion,'' {\em Sci Rep}, vol.~9, p.~4899, Mar 2019.

\bibitem{Arfkenbook}
G.~B. Arfken and H.~J. Weber, {\em Mathematical Methods for Physicists}.
\newblock San Diego: Academic Press, 2001.

\bibitem{Paulsen15}
J.~L. Paulsen, E.~{\"O}zarslan, M.~E. Komlosh, P.~J. Basser, and Y.-Q. Song,
  ``Detecting compartmental non-{G}aussian diffusion with symmetrized
  double-{PFG} {MRI},'' {\em NMR Biomed}, vol.~28, pp.~1550--6, Nov 2015.

\bibitem{Gradshteynbook}
I.~S. Gradshteyn and I.~M. Ryzhik, {\em Table of Integrals, Series, and
  Products}.
\newblock London: Academic Press, 6~ed., 2000.

\bibitem{Kroenke04}
C.~D. Kroenke, J.~J.~H. Ackerman, and D.~A. Yablonskiy, ``On the nature of the
  {NAA} diffusion attenuated {MR} signal in the central nervous system,'' {\em
  Magn Reson Med}, vol.~52, pp.~1052--9, Nov 2004.

\bibitem{Yablonskiy02}
D.~A. Yablonskiy, A.~L. Sukstanskii, J.~C. Leawoods, D.~S. Gierada, G.~L.
  Bretthorst, S.~S. Lefrak, J.~D. Cooper, and M.~S. Conradi, ``Quantitative in
  vivo assessment of lung microstructure at the alveolar level with
  hyperpolarized ${}^3${H}e diffusion {MRI},'' {\em Proc Natl Acad Sci U S A},
  vol.~99, pp.~3111--6, Mar 2002.

\bibitem{Anderson05}
A.~W. Anderson, ``Measurement of fiber orientation distributions using high
  angular resolution diffusion imaging,'' {\em Magn Reson Med}, vol.~54, no.~5,
  pp.~1194--1206, 2005.

\bibitem{McKinnon17}
E.~T. McKinnon, J.~H. Jensen, G.~R. Glenn, and J.~A. Helpern, ``Dependence on
  b-value of the direction-averaged diffusion-weighted imaging signal in
  brain,'' {\em Magn Reson Imaging}, vol.~36, pp.~121--127, Feb 2017.

\bibitem{Ozarslan18FiP}
E.~\"{O}zarslan, C.~Yolcu, M.~Herberthson, H.~Knutsson, and C.-F. Westin,
  ``Influence of the size and curvedness of neural projections on the
  orientationally averaged diffusion {MR} signal,'' {\em Front Phys}, vol.~6,
  p.~17, 2018.

\bibitem{Sen95}
P.~N. Sen, M.~D. H\"urlimann, and T.~M. {de Swiet}, ``Debye-{P}orod law of
  diffraction for diffusion in porous media,'' {\em Phys Rev B}, vol.~51,
  no.~1, pp.~601--604, 1995.

\bibitem{Shemesh09}
N.~Shemesh, E.~\"Ozarslan, P.~J. Basser, and Y.~Cohen, ``Measuring small
  compartmental dimensions with low-q angular double-{PGSE} {NMR}: the effect
  of experimental parameters on signal decay,'' {\em J Magn Reson}, vol.~198,
  no.~1, pp.~15--23, 2009.

\bibitem{OzarslanMRC11}
E.~\"Ozarslan, M.~Komlosh, M.~Lizak, F.~Horkay, and P.~Basser, ``Double pulsed
  field gradient (double-{PFG}) {MR} imaging ({MRI}) as a means to measure the
  size of plant cells.,'' {\em Magn Reson Chem}, vol.~49, pp.~S79--S84, Dec
  2011.

\bibitem{Komlosh13NI}
M.~E. Komlosh, E.~{\"O}zarslan, M.~J. Lizak, I.~Horkayne-Szakaly, R.~Z.
  Freidlin, F.~Horkay, and P.~J. Basser, ``Mapping average axon diameters in
  porcine spinal cord white matter and rat corpus callosum using {d-PFG}
  {MRI},'' {\em NeuroImage}, vol.~78, pp.~210--6, Sep 2013.

\bibitem{OzarslanBJ08}
E.~\"Ozarslan, U.~Nevo, and P.~J. Basser, ``Anisotropy induced by macroscopic
  boundaries: Surface-normal mapping using diffusion-weighted imaging,'' {\em
  Biophys J}, vol.~94, no.~7, pp.~2809--2818, 2008.

\bibitem{OzarslanJCP09}
E.~\"Ozarslan, N.~Shemesh, and P.~J. Basser, ``A general framework to quantify
  the effect of restricted diffusion on the {NMR} signal with applications to
  double pulsed field gradient {NMR} experiments.,'' {\em J Chem Phys},
  vol.~130, no.~10, p.~104702, 2009.

\bibitem{Moutal19tmi}
N.~Moutal, I.~I. Maximov, and D.~S. Grebenkov, ``Probing surface-to-volume
  ratio of an anisotropic medium by diffusion {NMR} with general gradient
  encoding,'' {\em IEEE Trans Med Imaging}, vol.~38, pp.~2507--2522, Nov 2019.

\bibitem{Jespersen12}
S.~N. Jespersen, ``Equivalence of double and single wave vector diffusion
  contrast at low diffusion weighting.,'' {\em NMR Biomed}, vol.~25,
  pp.~813--818, Jun 2012.

\bibitem{LiuOzarslan2019NBMreview}
C.~Liu and E.~\"{O}zarslan, ``Multimodal integration of diffusion {MRI} for
  better characterization of tissue biology,'' {\em NMR Biomed}, vol.~32,
  p.~e3939, Apr 2019.

\bibitem{Westin16_QTI}
C.~F. Westin, H.~Knutsson, O.~Pasternak, F.~Szczepankiewicz, E.~{\"O}zarslan,
  D.~van Westen, C.~Mattisson, M.~Bogren, L.~J. O'Donnell, M.~Kubicki,
  D.~Topgaard, and M.~Nilsson, ``Q-space trajectory imaging for
  multidimensional diffusion {MRI} of the human brain,'' {\em NeuroImage},
  vol.~135, pp.~345--62, Jul 2016.

\bibitem{Topgaard19DTDI}
D.~Topgaard, ``Diffusion tensor distribution imaging,'' {\em NMR Biomed},
  vol.~32, p.~e4066, 05 2019.

\bibitem{Ozarslan_JMR09}
E.~\"Ozarslan, ``Compartment shape anisotropy ({CSA}) revealed by double pulsed
  field gradient {MR}.,'' {\em J Magn Reson}, vol.~199, no.~1, pp.~56--67,
  2009.

\bibitem{Lawrenz10}
M.~Lawrenz, M.~A. Koch, and J.~Finsterbusch, ``A tensor model and measures of
  microscopic anisotropy for double-wave-vector diffusion-weighting experiments
  with long mixing times,'' {\em J Magn Reson}, vol.~202, pp.~43--56, Jan 2010.

\bibitem{Ianus17MAVAreview}
A.~Ianu\c{s}, N.~Shemesh, D.~C. Alexander, and I.~Drobnjak, ``Measuring
  microscopic anisotropy with diffusion magnetic resonance: from material
  science to biomedical imaging,'' in {\em Modeling, Analysis, and
  Visualization of Anisotropy} (T.~Schultz, E.~\"{O}zarslan, and I.~Hotz,
  eds.), Mathematics and Visualization, pp.~229--255, Springer International
  Publishing, 2017.

\bibitem{Jian07}
B.~Jian, B.~C. Vemuri, E.~\"Ozarslan, P.~R. Carney, and T.~H. Mareci, ``A novel
  tensor distribution model for the diffusion-weighted {MR} signal,'' {\em
  NeuroImage}, vol.~37, no.~1, pp.~164--176, 2007.

\bibitem{Leow09}
A.~D. Leow, S.~Zhu, L.~Zhan, K.~McMahon, G.~I. de~Zubicaray, M.~Meredith, M.~J.
  Wright, A.~W. Toga, and P.~M. Thompson, ``The tensor distribution function,''
  {\em Magn Reson Med}, vol.~61, pp.~205--14, Jan 2009.

\bibitem{Scherrer16}
B.~Scherrer, A.~Schwartzman, M.~Taquet, M.~Sahin, S.~P. Prabhu, and S.~K.
  Warfield, ``Characterizing brain tissue by assessment of the distribution of
  anisotropic microstructural environments in diffusion-compartment imaging
  ({DIAMOND}),'' {\em Magn Reson Med}, vol.~76, pp.~963--77, 09 2016.

\bibitem{Shakya17Dagstuhl}
S.~Shakya, N.~Batool, E.~\"Ozarslan, and H.~Knutsson, ``Multi-fiber
  reconstruction using probabilistic mixture models for diffusion {MRI}
  examinations of the brain,'' in {\em Modeling, Analysis, and Visualization of
  Anisotropy} (T.~Schultz, E.~\"Ozarslan, and I.~Hotz, eds.), pp.~283--308,
  Cham: Springer International Publishing, 2017.

\end{thebibliography}

\end{document}